\documentclass[a4paper, amsfonts, amssymb, amsmath, reprint, showkeys, nofootinbib, twoside, onecolumn]{revtex4-1}
\usepackage[english]{babel}
\usepackage[utf8]{inputenc}
\usepackage[colorinlistoftodos, color=green!40, prependcaption]{todonotes}
\usepackage{amsthm}
\usepackage{mathtools}
\usepackage{physics}
\usepackage{xcolor}
\usepackage{graphicx}
\usepackage[left=23mm,right=13mm,top=35mm,columnsep=15pt]{geometry} 
\usepackage{adjustbox}
\usepackage{placeins}
\usepackage[T1]{fontenc}
\usepackage{csquotes}
\usepackage[pdftex, pdftitle={Article}, pdfauthor={Author}]{hyperref} 
\begin{document}
\title{Remark on the Emergence of Color Superconductivity for Gauge Theories \\ in General Spacetime Dimensions from simple Holographic Models}

\author{Nguyen Hoang Vu}
    \email{vu@jinr.ru}
    \affiliation{The Bogoliubov Laboratory of Theoretical Physics, JINR,141980 Dubna, Moscow region, Russia}
    \affiliation{Institute of Physics, VAST, 10000, Hanoi, Vietnam}


\date{\today} 

\begin{abstract}
We generalize the concept of holography for the color superconductivity (CSC) phase by considering a $d$-dimensional Anti de Sitter (AdS) space instead of the traditional 6 dimensions. The corresponding dual field theory is an arbitrary confining gauge theory with $SU(N_c)$ symmetry, like quantum chromodynamics (QCD) CSC. We then use a holographic model based on Einstein-Maxwell gravity in $d$-dimensional AdS spacetime to study this phenomenon in both confinement and deconfinement phase, and we focus to study the confinement$-$deconfinement phase transition and the condition for the $N_c=2$ CSC phase with the $d=4$ case, one special case from \cite{Vu2024}. 
\end{abstract}


\maketitle

\section{Introduction}
In quantum chromodynamics (QCD) theory, the color superconductivity (CSC) phase is one of the interesting topics. It is the pairing of two quarks that is the difference colors in one Cooper pair condensate, the diquark. Unlike metallic superconductivity where the interaction between the electrons is repulsive and the electron pair attraction is caused by the interaction between electrons and phonons, the strong interaction between two quarks is attractive, and hence the quark Cooper pair formation (diquark formation) is more direct than electron Cooper pairs (but the diquarks only occur when the density is high enough). The diquark Cooper pair carry the net color charge, and therefore the condensation of the diquark spontaneously breaks the $SU(3)_C$ gauge symmetry of QCD and gives rise to the masses for the gluons via the Higgs mechanism. This phenomenon is related to color superconductivity (CSC), and this occurs at very high density (chemical potential) and low temperature. Thus, we believe that this phase exists in the inner cores of heavy neutron stars or quark stars \cite{fadafan2018,Fadafan2021,Kazuo2023} and hence we can probe this phase from the observation gravitational wave by LIGO or VIRGO \cite{LIGO2017} in the collision of neutron stars or quark stars \cite{Chen2025}. 

 One way to study the CSC phase is that we use the holographic principle or the AdS/CFT correspondence \cite{Maldacena1997,Witten98,Gubser1998} to approach. In this framework a weakly coupled gravity theory in the $d$-dimensional anti-de Sitter (AdS) spacetime corresponds to a strongly coupled conformal field theory (CFT) on the $(d-1)$-dimensional boundary of that spacetime. Within this frame work we can study one problem in the CFT with strong coupling constant by translating this to gravity problem at weak coupling constant via the holographic dictionary. In QCD, the CSC phase appears in high chemical potential and low temperature (below the QCD scale). To describe this by holographic QCD, we introduce an additional compact extra dimension on the boundary that corresponds to the QCD scale. This technique of geometrizing a physical effect has also been employed in classical physics \cite{phan2021curious}. As a result, for the AdS theory to be dual to our four$-$dimensional spacetime universe, the boundary becomes $R^{3,1}\times S^1$, and its bulk spacetime has six dimensions $AdS_6$ \cite{Basu.et.al.2011}. This differs from traditional holographic QCD \cite{Vu2020}, \cite{Arefeva2019} where we studied the five-dimensional bulk and the four dimensions found in holographic models of metallic superconductivity \cite{Horowitz2008,Hartnoll2008}.

In the first study of the holographic model for the CSC phase \cite{Basu.et.al.2011}, the authors considered Einstein-Maxwell gravity and the standard Maxwell interaction (i.e. $\mathcal{L}_{\text{Maxwell}}=-\frac{1}{4}F^2$) in the six$-$dimensional AdS spacetime. It is important to note that $\mathcal{L}_{\text{Maxwell}}$ differs from the Maxwell power-law holographic model studied in Cao \cite{Nam2022}. There, an AdS soliton corresponds to the confinement phase, while a Reissner$-$Nordström (RN) AdS black hole is dual to the deconfinement phase. The scalar hair (a complex scalar field) corresponds to the diquark operator in the boundary, called the $s-$ wave color superconductivity (because the color superconductivity can have p-wave and d-wave, analogy to the superconductivity in metallic and the high $T_c$ superconductor), which only appears when the chemical potential $\mu$ exceeds a critical value $\mu_c$ and, in the deconfinement phase, the temperature $T$ is below a critical temperature $T_c$, which depends on $\mu_c$ \cite{Basu.et.al.2011}. In these papers, the authors have proven that color superconductivity could not occur in the confinement phase and only appeared in the deconfinement phase, until Kazuo et al. \cite{Kazuo2019} have demonstrated that under Einstein-Maxwell gravity, the CSC phase is possible, but limited to cases with a single color $N_c = 1$. In more detail, the Breitenlohner-Freedman (BF) bound \cite{BF1,BF2} for the stability of another bulk scalar field with effective mass $m^2_{eff}$ (this mass square is the "natural" mass square in the equation of motion of the bulk scalar field) is broken only when $N_c < 1.89$ \cite{Kazuo2019}, therefore, we cannot study the CSC phase with $N_c \geq 2$ (because when $N_c\geq 2$, the bulk field which correspond to the diquark and the bulk field have the mass $m^2_{eff}$ are both stability hence we cannot have the condensate to the bulk field which is dual to the diquark). This problem may be solved by modifying the gravity framework or Maxwell's interaction law, as suggested in Cao \cite{Nam2021} and Cao \cite{Nam2022}. 

Here, there is an open question that is what happens if we generalize the conception of the CSC phase for an arbitrary $SU(N_c)$ and build a holographic model for this phase with an arbitrary bulk $d$ dimension $AdS_d$. In the previous project, \cite{Vu2024} we explored the holographic model for the CSC phase in the general case but without the confinement phase. In this project, we add the confinement phase (confined gauge theory) dual with the AdS soliton solution. We consider the confinement-deconfinement phase transition in the d-dimension by the critical chemical potential, and we will find the CSC phase in the confinement and deconfinement phases. In Section \ref{sec2}, we introduce the d-dimensional gravitational dual model of interest for the CSC phase transition. In Section \ref{sec3}, we analyze the CSC phase, deriving the conditions on $(d,N_c)$ for the formation of Cooper pairs on both the confinement and the deconfinement phases; we also discuss the critical chemical potential of the color superconductivity in a four-dimensional bulk and the condition of the trial function. Finally, in Section \ref{sec4}, we conclude with the main results and mention some open questions and interesting future directions.

\section{Holographic Model setup \label{sec2}}

First of all, we redefine the conception of the generalized color superconductivity phase. In this paper, it is one arbitrary Cooper pair condensate (we called the Cooper pair not the diquark because we don't consider the CSC in QCD) for one arbitrary $SU(N_c)$ gauge theory, which analogs the diquark Cooper pair in QCD color superconductivity. In analogy to the diquark condensate, the generalize Cooper pair also carry the net $SU(N_C)$ color charge hence it also spontaneously breaks the $SU(N_C)$ gauge symmetry. Now, we will consider the CSC phase in the case of an arbitrary confined gauge theory, like the QCD CSC case. From this assumption we have the confinement-deconfinement phase transition and we find the CSC phase in both the confinement and deconfinement phases. And, like the holographic QCD CSC we also have one compact dimension $y$, which corresponds to the scale, called confined gauge scale, in analogy of the QCD scale. Hence, the boundary of this case becomes $R^{d-2}\times S^1$ and the bulk $R^{d-1}\times S^1$, where $S^1$ is a circle with radius $R_y$ that corresponds to the scale \cite{Basu.et.al.2011}, instead of $R^{d-1}$ and $R^d$ in\cite{Vu2024}. The action for the $d$-dimensional Einstein-Maxwell gravity during the CSC phase transition is given by \cite{Emparan2014}:
 \begin{equation}
 \begin{split}
S=&\int d^dx\sqrt{-g}\Bigg[\mathcal{R}+\frac{(d-1)(d-2)}{L^2}\\
&-\frac{1}{4}F^2-|(\partial_{\mu}-iqA_{\mu})\psi|^2-m^2|\psi|^2 \Bigg] \ ,
 \label{EM_action}
 \end{split}
 \end{equation}
where $F_{\mu\nu}=\partial_{\mu}A_{\nu}-\partial_{\nu}A_{\mu}$ and the cosmological constant is determined by $\Lambda=-\frac{(d-1)(d-2)}{2L^2}$. We then set the AdS radius $L=1$ for convenience. Here, the $U(1)$ gauge field $A_{\mu}$ is dual to the general $SU(N_c)$ current analogous to the baryon number current in the CSC phase of QCD or the electric current in metallic superconductivity. The complex scalar field $\psi$ is dual to the boundary Cooper pair scalar field operator; specifically, in the holographic model for the QCD color superconductivity, it corresponds to the diquark Cooper pair scalar field operator (the $s-$wave CSC phase). The charge $q$ of this scalar field $\psi$ is associated with quantity in general, called general CSC charges, like the baryon number of the diquark in QCD color superconductivity, and its value is given by 
\begin{equation}
    q=\frac{2}{N_c} \ ,
\label{get_q}
\end{equation}
in which $N_c$ counts number of colors. 

Varying the action \eqref{EM_action} with respect to the vector and scalar fields we obtain the general equation of motion for he gauge field and the bulk scalar field as 
\begin{equation}
\label{general eom}
    \begin{split}
        \nabla_{\mu}F^{\mu\nu}-iq[\psi^*(\nabla^{\nu}-iqA^{\nu})\psi-\psi(\nabla^{\nu}+iqA^{\nu})\psi^*]&=0\\,
        (\nabla_{\nu}-iqA_{\nu})(\nabla^{\nu}+iqA^{\nu})\psi-m^2\psi&=0.
    \end{split}
\end{equation}

To further simplify our model from Eq. \eqref{EM_action}, we focus on $s-$wave CSC (may be the $p-$wave or $d-$wave CSC phase exist but we don't study these waves in this project), in which the vector field $A_\mu$ and complex scalar field $\psi$ follow the ansatz:
 \begin{equation}
     A_{\mu}dx^{\mu}=\phi(r)dt \ , \  \psi=\psi(r) \ ,
\label{ansatz}
 \end{equation}
where the variations are purely radial. The $s-$wave CSC phase appears from the condensation of the scalar field Cooper pairs (in $p-$wave and $d-$wave color superconductors, if these exist, the Cooper pairs are the vector fields) corresponding to the spontaneous breaking of the $U(1)$ gauge symmetry. Assuming that the charge is fixed, the condensation of the scalar field $\psi$ is controlled by the chemical potential, analogous to the baryon chemical potential of quarks in the QCD color superconductivity. At the critical chemical potential, the Cooper pair scalar condensation is created, and we have the dual bulk scalar field $\psi(r)=0$. Near the critical chemical potential, the value of the bulk scalar field $\psi\approx 0$ and we can neglect the back reaction of the bulk scalar field on the spacetime. Therefore, the back reaction of the matter field is only contributed by the $U(1)$ gauge field $A_{\mu}$. 

In this paper, we include the confinement phase; thus, we will find the CSC phase in both the confinement phase and the deconfinement phase. In the deconfinement phase that duals to black hole in the bulk, the phase transition of the CSC phase occurs at a critical temperature $T_c$ \cite{Basu.et.al.2011}, this temperature is associated with the critical chemical potential $\mu_c$ (similar to the QCD CSC phase in the deconfinement phase). The bulk scalar field that corresponds to the Cooper pair appears when the chemical potential $\mu>\mu_c$ or the temperature $T<T_c$, called the scalar hair of a black hole. In the holographic dictionary, because there is one scalar hair, the spacetime geometry dual to this phase is described by the Reissner-Nordström (RN) planar black hole solution, with the metric given by the following ansatz:
 \begin{equation}
  \label{black hole solution}
 ds^2=r^2 \Big[-f(r)dt^2+h_{ij}dx^idx^j +dy^2\Big]+\frac{dr^2}{r^2f(r)} \ ,
 \end{equation}
 where $h_{ij}dx^idx^j=dx_1^2+...+dx_{d-3}^2$ is the line element of the $(d-3)$-dimension hypersurface and the direction $y$ is compacted with the radius $R_y$. The event horizon radius $r_+$ satisfies 
 \begin{equation}
      f(r_+)=0 \ .
\label{black_hor}
 \end{equation}
 In the holographic dictionary, the temperature of the boundary field theory is associated with the Hawking temperature of this $d$-dimensional RN planar AdS black hole, i.e. 
 \begin{equation}
 T=T_H\equiv\frac{r_+^2f'(r_+)}{4\pi} \ .
 \label{Hawking}
 \end{equation}

Using the ansatz Eq.\eqref{ansatz}, from Eq.\eqref{general eom}, we can obtain the classical equations of motion for the temporal component of the vector field $\phi$ and the complex scalar field $\psi$ to be:
  \begin{equation}
  \label{eomdeconfinement}
\begin{split}
&\phi''(r)+\frac{d-2}{r}\phi'(r)-\frac{2q^2\psi^2(r)}{r^2f(r)}\phi(r)=0 \ , 
\\
&\psi''(r)+\left[\frac{f'(r)}{f(r)}+\frac{d}{r}\right]\psi'(r)+\frac{1}{r^2f(r)}\left[\frac{q^2\phi^2(r)}{r^2f(r)}-m^2\right]\psi(r)=0 \ ,
\end{split}
\end{equation}
in which the blackening function $f(r)$ is given by \cite{Nam2021},\cite{Nam2019}:
 \begin{equation}
 f(r)=1-\left(1+\frac{3\mu^2}{8r_+^2}\right)\left(\frac{r_+}{r}\right)^{d-1}+\frac{3\mu^2r_+^d}{8r^{d+2}}
 \label{blackening}
 \end{equation}
 To check the blackening function, in the case of the spacetime of $d=6$, this expression becomes:
 $$f(r)=1-\left(1+\frac{3\mu^2}{8r^2_+}\right)\left(\frac{r_+}{r}\right)^5+\frac{3\mu^2r_+^6}{8r^8} \ , $$ 
 which is equivalent to the holographic model for the QCD CSC in the deconfinement region \cite{Kazuo2019}. 

In the deconfinement phase, the temperature $T$ of the color superconductivity phase living on the boundary corresponds to the Hawking temperature $T_H$ observed in the bulk, as mentioned in Eq. \eqref{Hawking}. Hence:
 \begin{equation}
 T=\frac{r_+^2f'(r_+)}{4\pi}=\frac{1}{4\pi}\left[(d-1)r_+-\frac{9\mu^2}{8r_+}\right] \ .
 \end{equation}
 From the physical condition that these thermal temperatures cannot be negative, we have the constraint for the chemical potential $\mu$:
  \begin{equation}
 \frac{\mu^2}{r_+^2}\leq \frac{8(d-1)}{9} \ .
 \label{chempot_constraint}
 \end{equation}
 
 Near the boundary $(r\rightarrow\infty)$, from the equations of motion Eq. \eqref{eomdeconfinement}, we have the asymptotic forms for the matter fields:
\begin{equation}\label{asymptotic form}
\begin{split}
\phi(r)&= \mu -\frac{\rho}{r^{d-3}} \ , 
\\
\psi(r)&=\frac{J_C}{r^{\Delta_-}}+\frac{C}{r^{\Delta_+}} \ ,
\end{split}
\end{equation}
where $\mu,\rho,J_C$, and $C$ are regarded as the chemical potential, charge density, source, and the condensates vacuum expected value (VEV) of the Cooper pair operator dual to $\psi$ (like the diquark Cooper pair in QCD), respectively.  In this case, the conformal dimensions $\Delta_{\pm}$ read:
\begin{equation}
\Delta_{\pm}=\frac{1}{2}\left[(d-1)\pm\sqrt{(d-1)^2+4m^2}\right] \ ,
\end{equation}
and the BF bound -- as follows from \cite{BF1,BF2} -- is:
\begin{equation}
\label{BF bound}
    m^2\geq-\frac{(d-1)^2}{4} \ .
\end{equation}
This is the stability condition for the field $\psi$. To simplify, we set $\Delta_-=1$ \cite{Nam2019}, which leads to:
\begin{equation}
    m^2=2-d  \ ,
\label{mass_sq}
\end{equation} 
and therefore obtain $\Delta_+=d-2$. Note that in this setting, two components of $\psi$ are normalizable modes, and $m^2$ always satisfies Eq. \eqref{BF bound}. Thus, Eq. \eqref{asymptotic form} becomes:
\begin{equation}
    \psi(r)=\frac{J_C}{r}+\frac{C}{r^{d-2}} \ .
\label{psi_func}
\end{equation}


From \cite{Vu2024} we have the boundary condition at
\begin{equation}
    \begin{split}
 &\phi(r_+)=0 \ , \\
        &\psi(r_+)=r_+^2\frac{f'(r_+)\psi'(r_+)}{m^2} \ .
    \end{split}
\label{horizon_condition}
\end{equation}
Now we consider the confinement phase. In the confinement phase the temperature $T=0$, the spacetime geometry dual to the confinement phase the AdS soliton solution \cite{Natsuume2015} in $d-$dimension and we have 
\begin{equation}
\label{AdS soliton}
    ds^2=r^2(\gamma_{\mu\nu}dx^{\mu}dx^{\nu}+dx_{d-3}^2+f(r)dy^2)+\frac{dr^2}{r^2f(r)},
\end{equation}
with 
\begin{equation}
    f(r)=1-\left(\frac{r_0}{r}\right)^{d-1}
\end{equation}
$r_0=\frac{2}{(d-1)R_y}$ is the radius which analogy with the event horizon. From this ansatz and Eq.\eqref{general eom} we obtain the equation of motion for the confinement phase
\begin{equation}
\label{confinement eom}
\begin{split}
    &\phi''(r)+\left[\frac{d-2}{r}+\frac{f'(r)}{f(r)}\right]\phi'(r)-\frac{2q^2\psi^2(r)}{r^2f(r)}\phi(r)=0\\
    &\psi''(r)+\left[\frac{d}{r}+\frac{f'(r)}{f(r)}\right]\psi'(r)+\frac{1}{r^2f(r)}\left[\frac{q^2\phi^2(r)}{r^2}-m^2\right]\psi(r)=0
\end{split}
\end{equation}
The boundary condition at $r_0$ \cite{Kazuo2019}, \cite{Nam2021}
\begin{equation}
\begin{split}
 \phi'(r_0)&=\frac{2q^2\psi^2(r_0)}{r_0^2f'(r_0)}\phi(r_0)\\
 \psi'(r_0)&=-\frac{1}{r_0^2f'(r_0)}\left(\frac{q^2\phi^2(r_0)}{r_0^2}-m^2\right)\psi(r_0).
\end{split}
\end{equation}
And replace the value of $f(r)$ in confinement phase, we get
\begin{equation}
\begin{split}
 \phi'(r_0)&=\frac{2q^2\psi^2(r_0)}{(d-1)r_0}\phi(r_0)\\
 \psi'(r_0)&=-\frac{1}{(d-1)r_0}\left(\frac{q^2\phi^2(r_0)}{r_0^2}-m^2\right)\psi(r_0)
\end{split}
\end{equation}

\section{The emergence of CSC phase with multiple colors \label{sec3}}

When the chemical potential $\mu$ exceeds the critical value $\mu_c$, Cooper pair condensation occurs. Near the critical chemical potential (when $\mu > \mu_c$ but still close to $\mu_c$), $\psi$ should be approximately $0$, so its back reaction can be neglected. Therefore, the bulk configuration is approximately determined by:
 \begin{equation}
 S=\int d^dx\sqrt{-g}\left(\mathcal{R}-2\Lambda-\frac{1}{4}F^2\right) \ .
 \end{equation}
The solution of the gauge field in this case is simply:
\begin{equation}
    \phi(r)=\mu\left[1-\left(\frac{r_+}{r}\right)^{d-3}\right] \ ,
\label{simple_sol}
\end{equation}
in the deconfinement phase duality with the planar RN black hole and
\begin{equation}
    \phi(r)=\mu,
\end{equation}
in the confinement phase duality to AdS soliton solution,
which is consistent with Eq. \eqref{asymptotic form} and Eq. \eqref{horizon_condition}.

To probe the confinement$-$deconfinement phase transition, we need to compute the Euclidean action of the bulk. The general Euclidean action is given by
\begin{equation}
    S^E=-\int d^dx\sqrt{-g}\mathcal{L}
\end{equation}
Using the result of \cite{Olea2011} for $d>4$ and $\alpha=0$ we obtain
\begin{equation}
\label{even Euclidean action}
\begin{split}
     S^E_{2k}&=\beta Vol(\Gamma_{d-2})[(r^2f)'r^{d-2}|^{\infty}_{r_+}-(r^2f)'(r^2f)^{k-1}|^{\infty}-r^{d-2}\phi\phi'|^{\infty}_{r_+} ] 
\end{split}
\end{equation}
with $d$ is even, $d=2k$ and with $d$ odd we have
\begin{equation}
\label{odd Euclidean action}
\begin{split}
    &S^E_{2k+1}=\beta Vol(\Gamma_{2k-1})r^{2k-1}(r^2f)'|^{\infty}_{r_+}\\
    &+\beta Vol(\Gamma_{2k-1})kc_{2k}(2k-1)!\\
    &\times[r^{2k-1}(r^2f)'\int_0^1dt(\frac{-r^4f^2}{r^2}+t^2)^{k-1}+2\int^1_0dtt(r^4f^2-\frac{r(r^4f^2)'}{2})(t^2(r^2-r^4f^2))^{k-1}]|^{\infty}\\
    &-\beta Vol(\Gamma_{d-2})r^{d-2}\phi\phi'|^{\infty}_{r_+}
\end{split}    
\end{equation}
with $t$ is one parameter to express the boundary term becomes the polynomial in \cite{Olea2011} and 
\begin{equation}
  c_{2k}=-\frac{2(-1)^{k-1}}{k(2k-1)!\beta(k,\frac{1}{2})}.  
\end{equation}

From \eqref{even Euclidean action} and \eqref{odd Euclidean action} we obtain the free energy for the AdS black hole and for the AdS soliton when $d$ even. 
\begin{equation}
\label{even free energy}
\begin{split}
     \Omega_{BH{2k}}&= Vol(\Gamma_{d-2})[(r^2f)'r^{d-2}|^{\infty}_{r_+}-(r^2f)'(r^2f)^{k-1}|^{\infty}-r^{d-2}\phi\phi'|^{\infty}_{r_+} ] \\
    &=[-r_+^{d-1}-\frac{5d-27}{8}\mu^2r_+^{d-3}]Vol(\Gamma_{d-2}),
\end{split}
\end{equation}
with the AdS black hole and with the AdS soliton we have
\begin{equation}
    \Omega_{soliton{2k}}=-r_0^{d-1}Vol(\Gamma_{d-2}),
\end{equation}
for $d=2k$.

When $d$ odd, $d=2k+1$ we have the free energy
\begin{equation}
\label{odd free energy}
\begin{split}
    &\Omega_{{2k+1}}= Vol(\Gamma_{2k-1})r^{2k-1}(r^2f)'|^{\infty}_{r_+}\\
    &+Vol(\Gamma_{2k-1})kc_{2k}(2k-1)!\\
    &\times[r^{2k-1}(r^2f)'\int_0^1dt(-r^2f^2+t^2)^{k-1}
    +2\int^1_0dtt(r^4f^2-\frac{r(r^4f^2)'}{2})(t^2(r^2-r^4f^2))^{k-1}]|^{\infty}\\
    &-Vol(\Gamma_{d-2})r^{d-2}\phi\phi'|^{\infty}_{r_+}.
\end{split}    
\end{equation}
Here 
\begin{equation}
    \begin{split}
        &f(r)=1-\left(1+\frac{3\mu^2}{8r_+^2}\right)\left(\frac{r_+}{r}\right)^{d-1}+\frac{3\mu^2r_+^d}{8r^{d+2}},\\
        &\phi(r)=\phi(r)=\mu\left[1-\left(\frac{r_+}{r}\right)^{d-3}\right],
    \end{split}
    \label{AdS bh}
\end{equation}
with the AdS black hole and
\begin{equation}
    \begin{split}
        &f(r)=1-\left(\frac{r_0}{r}\right)^{d-1},\\
        &\phi(r)=\mu,
    \end{split}
\end{equation}
with the AdS solition case. From the result of \cite{Vu2024} and \cite{Nam2021}, we only consider the free energy in the case of even dimension because in the case of $d=4$ we can study the multicolored CSC phase in the case of the non-confinement phase \cite{Vu2024} and in the case of $d=6$ we have the case of QCD color superconductivity. In the case of an odd dimension, we will evaluate the maximum of the number of colors $N_{cmax}$ by estimate. 

These formulas cannot apply for the case $d=4$. In $d=4$ case, the Euclidean action becomes
\begin{equation}
    S^E=S_{grav}+S^E_{matter}+S^E_{CT}
\end{equation}
The gravity action $S_{grav}$ is given by
\begin{equation}
    S_{grav}=S_{bulk grav}+S_{G.H.Y}
\end{equation}
 We have
 \begin{equation}
 \label{bulk grav}
    \begin{split}
         S_{bulk grav}&=-\int dX^4\sqrt{-g}(R-2\Lambda)=2XR_y\beta(4-1)\frac{1}{3}r^3|^{\infty}_{r_+}+\int dX^4\frac{9\mu^2}{4}\frac{r^4_+}{r^6}\\
         &=2XR_y\beta r^3|^{\infty}_{r_+}-XR_y\beta\frac{3\mu^2}{4}r_+^4\frac{1}{r^3}|^{\infty}_{r_+}=2XR_y\beta r^3|^{\infty}_{r_+}+XR_y\beta\frac{3\mu^2r_+}{4}
    \end{split}
 \end{equation}
 The Gibbons-Hawking-York term
 \begin{equation}
     S_{GHY}=-2\int dx^3\sqrt{-h}K
 \end{equation}
 Where $K$ is the extrinsic curvature and $h=det h_{ab}$ with $h_{ab}$ the metric in $d-1$ dimension, without $r$. Hence $h_{ab}=(-r^2f(r),r^2,r^2)$ and $h^{ab}=(\frac{1}{r^2f(r)},\frac{1}{r^2},\frac{1}{r^2})$, and the normal vector is defined by
\begin{equation}
    n^{\mu}=\frac{1}{\sqrt{g_{rr}}}\left(\frac{\partial}{\partial r}\right)^{\mu}=\frac{\delta_r^{\mu}}{\sqrt{g_{rr}}}
\end{equation}
The non zero component of the normal vector is
\begin{equation}
    n^r=r\sqrt{f(r)}
\end{equation}
 We have
 \begin{equation}
     K=\frac{1}{2}h^{ab}n^r\partial_rh_{ab}=\frac{1}{2}\left(\frac{rf'(r)}{\sqrt{f(r)}}+2\sqrt{f(r)}+4\sqrt{f(r)}\right)
 \end{equation}
 and
 \begin{equation}
     \sqrt{-h}=\sqrt{-(-r^2f(r).r^2.r^2)}=r^3\sqrt{f(r)}
 \end{equation}
 Hence the G.H.Y term
 \begin{equation}
 \label{GHY}
     S_{GHY}=\beta R_yX(-6r^3f(r)-r^4f'(r))|^\infty
 \end{equation}
 The matter part $S^E_{matter}$ is given by
 \begin{equation}
     S^E_{matter}=S^E_{bulk matter}+S_{bnd,F}
 \end{equation}
We have
\begin{equation}
    \begin{split}
        S^E_{bulk matter}&=-XR_y\beta\int dr\sqrt{-g}\left(-\frac{1}{4}F^2\right)\\
        &=-XR_y\beta\int dr\sqrt{-g}\left(-\frac{1}{2}g^{00}g^{rr}\phi'^2\right)\\
        &=-XR_y\beta\frac{1}{2}r^2\phi(r)\phi'(r)|^{\infty}_{r_+}\\
        &=-XR_y\beta\frac{1}{2}\mu^2r_+
    \end{split}
\end{equation}
and
\begin{equation}
    \begin{split}
        S_{bnd,F}&=XR_y\beta\frac{1}{2}\sqrt{-h}n_aF^{ab}A_b\\
        &=XR_y\beta\frac{1}{2}\sqrt{-h}n_rF^{r0}A_0\\
        &=-XR_y\beta\frac{1}{2}r^3\sqrt{f(r)}\frac{1}{r\sqrt{f(r)}}\phi'(r)\phi(r)|^{\infty}\\
        &=-XR_y\beta\frac{1}{2}r^2\phi\phi'|^{\infty}\\
        &=-XR_y\beta\frac{1}{2}\mu^2r_+
    \end{split}
\end{equation}
We have
\begin{equation}
\label{matter}
    S^E{matter}=-XR_y\beta\mu^2r_+
\end{equation}
And finally the counter-term action is given by
\begin{equation}
\label{counter term}
    \begin{split}
        S^E_{CT}&=2(d-2)\sqrt{-h}\beta XR_y\\
        &=4\beta XR_yr^3\sqrt{f(r)}
    \end{split}
\end{equation}
From \eqref{counter term}, \eqref{bulk grav}, \eqref{GHY}, \eqref{matter}, we obtain the Euclidean action for $d=4$ case 
\begin{equation}
\label{4d bh}
    S^E_{4dbh}=-r^3_+\left(1-\frac{\mu^2}{8r_+^2}\right)\beta XR_y,
\end{equation}
and with the 4d AdS soliton case
\begin{equation}
\label{4d soliton}
    S^E_{4d soliton}=-r_0^3\beta XR_y
\end{equation}
Hence, at $4d$ we have $\Omega_{4dbh}=-r^3_+\left(1-\frac{\mu^2}{8r_+^2}\right)XR_y$ and $\Omega_{4d soliton}=-r^3_0XR_y$

The confinement-deconfinemnt phase transition in this paper corresponds to the Hawking-Pgae transition from the AdS soliton to the AdS black hole. The critical point is probed by solving equation $\Omega_{BH}=\Omega_{soliton}$ when $T=0$ (because the confinement-deconfinement phase transition in this model occurs when $T=0$ and $\mu>0$). We have the critical point of the confinement-deconfinement phase transition at $\mu_{cd(4)}=\sqrt{8/3}(\frac{3}{2})^{1/3}>\sqrt{8/3}$, we see that the critical chemical potential of the phase transition is greater than the maximum of the chemical potential in the case $d=4$. Hence, in the case $d=4$, this model does not have a phase transition of confinement and deconfinement. Hence, in $d=4$ we have 2 cases: the case without confinement and the case with full confinement for confined gauge theory (no phase transition).

With $d=6$ we find that the confinement$-$deconfinement phase transition ($c-d$ phase transition) occurs in $\mu_{cd(6)}=1.73$ \cite{Basu.et.al.2011}, \cite{Nam2021},\cite{Kazuo2019}. We also obtain the results of the confinement$-$deconfinement critical chemical potential $\mu_{cd}$ when $d=8$ is $\mu_{cd(8)}=1.76839$ and when $d=10$ it is $\mu_{cd(10)}=1.98696$. But with $d=4$ we have 

\subsection{CSC phase in deconfinement phase}

In \cite{Vu2024} we have proven that only with $d=4$ we have the color superconductivity phase with $N_c=2$ without the confinement phase by Einstein$-$Maxwell gravity. and with $d=2$, or $d=3$ the CSC phase does not exist. In this paper, it corresponds to the $Nc=2$ CSC phase that occurs in the deconfinement phase in $d=4$. Now we will quick review this proof. 

From the equation of motion for the deconfinement phase \eqref{eomdeconfinement}, we introduce the effective mass
\begin{equation}
\label{meff}
    m^2_{eff}=m^2-\Delta m^2=m^2-\frac{q^2\phi^2(r)}{r^2f(r)}
\end{equation}
From \cite{Vu2024} we have the instability condition of the effecitve mass that
\begin{equation}
\label{instability condition}
    m^2_{eff}<-\frac{(d-1)^2}{4}
\end{equation}

After some manipulation (detail in \cite{Vu2024}) we obtain
\begin{equation}
    N_c<\frac{4\sqrt{F_{max}(d,\hat{\mu},z)}}{d-3}
\end{equation}
In this formula $\hat{\mu}=\frac{\mu}{r_+}$, $z=\frac{r}{r_+}$ and the function $F(d,\hat{\mu},z)=\frac{\hat{\mu}^2z^2(1-z^{(d-3)})^2}{1-\left(1+\frac{3\mu^2}{8}\right)z^{d-1}+\frac{3\mu^2z^{d+2}}{8}}$

To study this function we introduce $F_{\max}(d,\hat{\mu}) \equiv \max_{z\in [0,1]} F(d,\hat{\mu},z)$, $\tilde{\mu}=\frac{3\hat{\mu}}{\sqrt{8(d-1)}}$ and
\begin{equation}
\label{G_man}
G(d,\tilde{\mu})\equiv\frac{4\sqrt{F_{max}(d,\tilde{\mu})}}{d-3} ,  
\end{equation} 
we have the instability condition becomes
\begin{equation}
    N_c<G(d,\tilde{\mu})
\end{equation}

\begin{figure*}[!htbp]
\includegraphics[width=\textwidth]{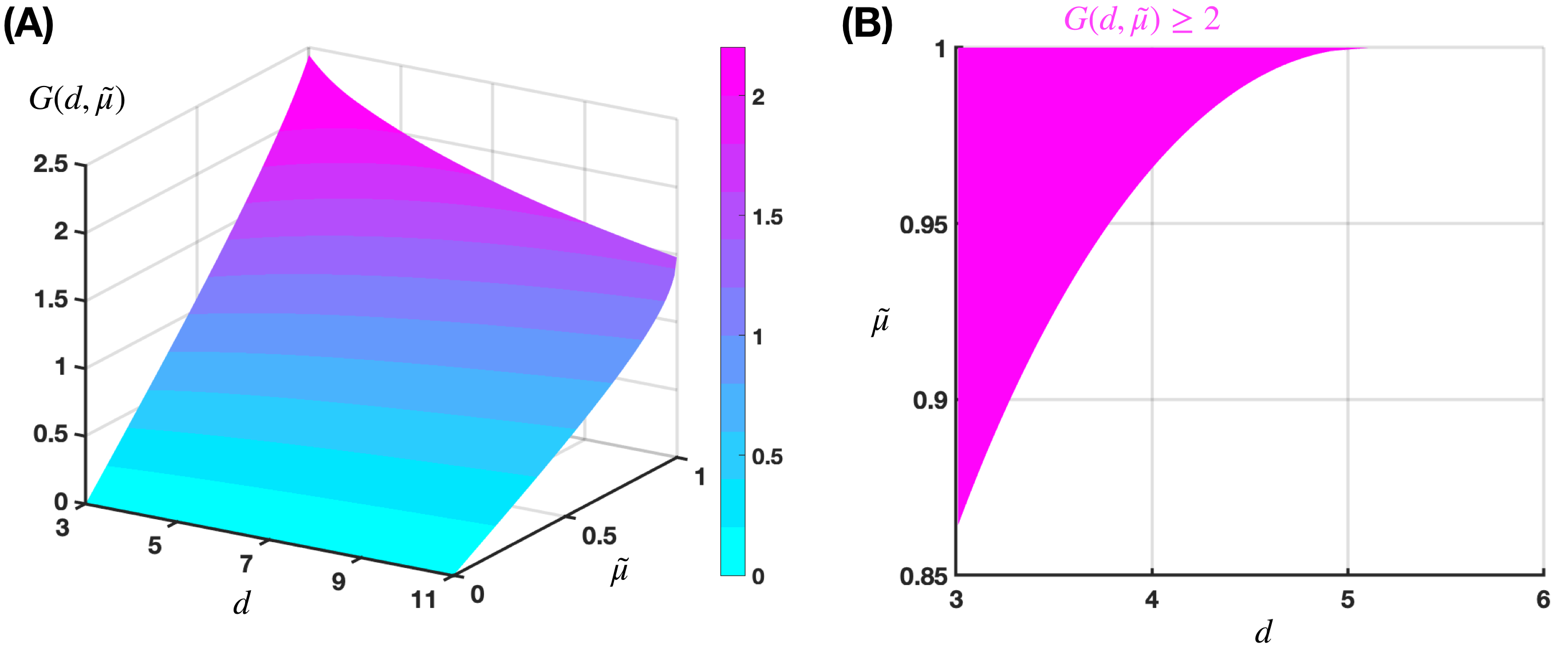}%
\caption{\textbf{Our numerical investigation for $G(d,\tilde{\mu})$ in Eq. \eqref{G_man}.} This calculation was done using MatLab R2023a \cite{MATLAB}. \textbf{(A)} The surface function $G(d,\tilde{\mu})$ inside the region of interests i.e. $(d,\tilde{\mu}) 
\in (3,11] \times [0,1]$. \textbf{(B)} We zoom into the small corner where $G(d,\tilde{\mu})>2$ can be realized. Image taken from \cite{Vu2024}.}
\label{fig01}
\end{figure*}

From Fig.\ref{fig01} we probe that only with the $d=4$ Einstein-Maxwell gravity can study the CSC phase with $N_c=2$. Now we focus on the case $d=4$ and solve by estimate the equation of motion \eqref{eomdeconfinement} near the critical point to find the critical chemical potential in non-confinement case by Sturm$-$Liouville method

From \cite{Kazuo2019}, above the critical chemical potential when the CSC phase appears, there is a scalar field solution with $J_c=0$ (because the broken symmetry is spontaneous) and $C\ne 0$. Hence the scalar field near boundary in this case becomes
\begin{equation}
\label{psi near boundary}
    \psi(r)=\frac{C}{r^{d-2}}
\end{equation}
Use the variable $z$ we have the second equation of \eqref{eomdeconfinement} and $r_+=1$
\begin{equation}
\label{eom z variable}
\begin{split}
   &\psi''(z)+\left[\frac{f'(z)}{f(z)}-\frac{d-2}{z}\right]\psi'(z)+\left[\frac{q^2\phi^2(z)}{f^2(z)}-\frac{m^2}{f(z)z^2}\right]\psi(z)=0
\end{split}
\end{equation}
From the blackening function \eqref{blackening} we obtain:
\begin{equation}
    \frac{f'(z)}{f(z)}=\frac{-(d-1)\left(1+\frac{3\mu^2}{8}\right)z^{d-2}+\frac{3\mu^2}{8}(d+2)z^{d+1}}{1-\left(1+\frac{3\mu^2}{8}\right)z^{d-1}+\frac{3\mu^2}{8}z^{d+2}}
\end{equation}
From the boundary condition \eqref{psi near boundary} we have the form of $\psi(z)$:
\begin{equation}
    \psi=Cz^{d-2}H(z)
\end{equation}
the function $H(z)$ is trial function and it satisfy the boundary condition $H(0)=1$ and $H'(0)=0$. And because this solution is close to the critical chemical potential (above but near), we can consider $\mu\approx\mu_c$. We obtain the equation for $H(z)$
\begin{equation}
\begin{split}
     &H''(z)+\frac{f'(z)z+(d-2)f(z)}{zf(z)}H'(z)+\frac{(d-2)[f'(z)z-f(z)]-m^2}{z^2f(z)}H(z)+q^2\mu^2_{c}\frac{(1-z^{d-3})^2}{f^2(z)}H(z)=0
\end{split}
\end{equation}
We rewrite this equation
\begin{equation}
    H''(z)+p(z)H'(z)+q(z)H(z)+\lambda^2 w(z)\xi^2(z)H(z)=0
\end{equation}
where $\lambda^2=q^2\mu_c^2$ and
\begin{equation}
    \begin{split}
        &p(z)=\frac{f'(z)}{f(z)}+\frac{d-2}{z}\\
        &q(z)=\frac{d-2}{z}\left(\frac{f'(z)}{f(z)}-\frac{1}{z}\right)-\frac{m^2}{f(z)z^2}\\
        &=\frac{(d-2)[-1-d\left(1+\frac{3\mu_c^2}{8}\right)z^{d-1}+\frac{3\mu_c^2}{8}(d+1)z^{d+2}]}{[1-\left(1+\frac{3\mu_c^2}{8}\right)z^{d-1}+\frac{3\mu_c^2}{8}z^{d+2}]^2}\\
        &-\frac{m^2}{[1-\left(1+\frac{3\mu_c^2}{8}\right)z^{d-1}+\frac{3\mu_c^2}{8}z^{d+2}]z^2}\\
       & w(z)=\frac{1}{f^2(z)}=\frac{1}{[1-\left(1+\frac{3\mu_c^2}{8}\right)z^{d-1}+\frac{3\mu_c^2}{8}z^{d+2}]^2}\\
       &\xi^2(z)=(1-z^{d-3})^2
    \end{split}
\end{equation}
It's written in form of the Sturm$-$Liuoville equation
\begin{equation}
\label{Sturm-Liouville equation}
    [T(z)H'(z)]'-Q(z)H(z)+\lambda^2P(z)H(z)=0
\end{equation}
where
\begin{equation}
\label{S-L coeff}
    \begin{split}
        Q(z)&=-T(z)q(z)\\
        P(z)&=T(z)w(z)\xi^2(z)\\
        T(z)&=e^{\int p(z)dz}=f(z)z^{d-2}\\
        &=[1-\left(1+\frac{3\mu_c^2}{8}\right)z^{d-1}+\frac{3\mu_c^2}{8}z^{d+2}]z^{d-2}
    \end{split}
\end{equation}
From the Sturm$-$Liouville equation, the eigenvalue $\lambda^2$ in \eqref{Sturm-Liouville equation} is obtained by minimizing the following expression
\begin{equation}
\label{eigenvalue}
    \lambda^2=\frac{\int_0^1 T(z)H'^2(z)dz+\int^1_0 Q(z)H(z)^2dz}{\int^1_0 P(z)H^2(z)dz}
\end{equation}
where the trial function $H(z)$ is chosen as $H(z)=1-az^2$ \cite{Siopsis2010} \cite{Nam2019}.
\begin{figure*}[!htbp]
\includegraphics[width=\textwidth]{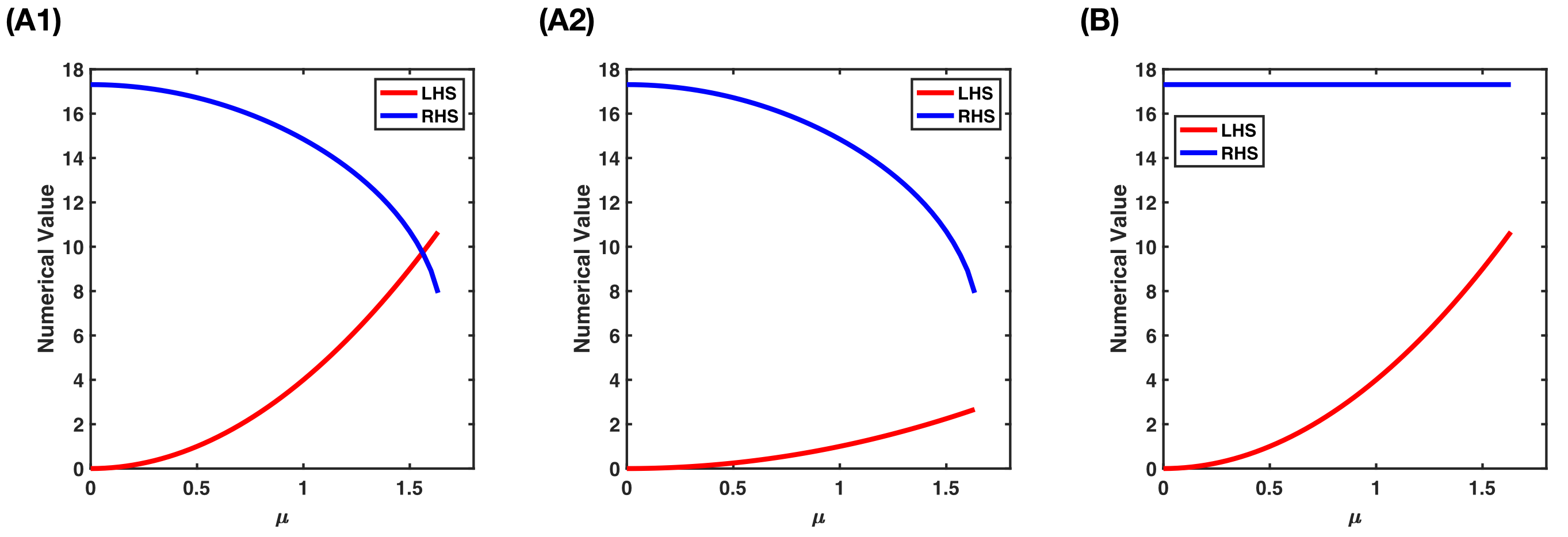}%
\caption{\textbf{Our numerical investigation for the eigenvalue estimation of the Sturm-Liouville equation.} These calculations was done using MatLab R2023a \cite{MATLAB}, using the test function $H(z)=1-az^2$. \textbf{(A1)} The case without confinement phase with $N_c=1$, corresponding to Eq. \eqref{eigenvalue}. \textbf{(A2)}  The case without confinement phase with $N_c=2$, corresponding to Eq. \eqref{eigenvalue}. \textbf{(B)} Case full confinement phase with $N_c=1$, corresponding to Eq. \eqref{S-L confinement}.}
\label{fig02}
\end{figure*}

With $Nc=1$ we can see that the critical chemical potential exists (see Fig. \ref{fig02}A1). Hence, in non-confinement case with $N_c=1$, the Einstein$-$Maxwell gravity can study color superconductivity when $d=4$.

But when we solve this equation with $H(z)=1-az^2$ and $N_c=2$ with $d=4$ we don't see the critical point (see Fig. \ref{fig02} A2), so in the bulk of dimensions $4$, we cannot study the color superconductivity phase via holography with $N_c=2$ by the simplest trial function $H(z)=1-az^2$

However, because the function $H(z)=1-az^2$ is only the simplest form of the trial function, therefore we cannot yet conclude that when $N_c=2$ in $d=4$  Einstein-Maxwell gravity only studies the CSC phase with $N_c=1$ in the non-confinement case because there are many forms of the trial function $H(z)$. Now, we need to try the RHS of \eqref{eigenvalue} with other forms of $H(z)$ that satisfies the boundary condition $H(0)=1$ and $H'(0)=0$. In $d=4$, if we find one continuous function $H(z)$ that satisfies the RHS of \eqref{eigenvalue} greater than $0$ and less than $8/3$ when $\mu=\sqrt{8/3}$, we will confirm that $\mu_c$ for CSC with $N_c=2$ . ($H(z)$ is still satisfy $H(0)=1$, and $H'(0)=0$). 
\begin{equation}
\label{condition 4d without confinement}
    \begin{split}
        &\text{RHS}_{f(z)=1-2z^3+z^6}<8/3\\
        &\text{RHS}_{f(z)=1-z^3}>0
    \end{split}
\end{equation}

These are because $f(z)=1-z^3$ when $\mu=0$ and $f(z)=1-2z^3+z^6$ when $\mu=\sqrt{8/3}$. The two conditions in \eqref{condition 4d without confinement} and $H(0)=1$, $H'(0)=0$ are all conditions for the trial function $H(z)$ to obtain the case of CSC phase without confinement phase. If there is a $H(z)$ that satisfies these conditions, we will confirm that the CSC with $N_c=2$ exists when $d=4$. If not we have only CSC phase with $N_c=1$ for all dimension in Einstein$-$Maxwell gravity. The proof of whether the function $H(z)$ exists or not is to be studied in the future. 

In the $4d$ non-confinement case, (because the scale not exist, we replace $XR_y$ in \eqref{4d bh} by the $V_2$), when we have a Cooper pair condensate near a critical chemical potential ($\mu>\mu_c$ but near), the free energy  in this case is as follows:
\begin{equation}
    \Omega_{4dhair}=\left(-r^3_+\left(1-\frac{\mu^2}{8r^2_+}\right)+\int^{\infty}_{r_+}\frac{q^2r^2\phi^2\psi^2}{f(r)}dr\right)\beta V_2
\end{equation}
with $N_c=1$ and use the variable $z$ and $r_+=1$ we have the free energy density:
\begin{equation}
\Omega_{4dhair N=1 dens}= 1-\frac{\mu_c^2}{8}+4\int^1_0\frac{\mu_c^2(1-z)^2C^2(1-az^2)^2}{f(z)}dz,
\end{equation}
and with $N_c =2$ (if it exist)
\begin{equation}
\Omega_{4dhair N=2 dens}= 1-\frac{\mu_c^2}{8}+\int^1_0\frac{\mu_c^2(1-z)^2C^2H(z)^2}{f(z)}dz.
\end{equation}
With the case $d>4$ and even, we have the free energy density with hair:
\begin{equation}
    \Omega_{even dim dens}=-[r_+^{d-1}+\frac{5d-27}{8}\mu^2r_+^{d-3}]+\int^{\infty}_{r_+}\frac{q^2r^2\phi^2\psi^2}{f(r)}dr.
\end{equation}

\subsection{CSC phase in confinement phase}

We now consider for the confinement phase case from \eqref{confinement eom} we also have the effective mass is 
\begin{equation}\label{conf effective mass}
    m^2_{eff}=m^2-\frac{q^2\phi^2(r)}{r^2}=m^2-\frac{q^2\phi^2(r)}{r_0^2}\left(\frac{r_0}{r}\right)^2
\end{equation}
From the BF bound breaking condition (with $r_0=1$) we have
\begin{equation} 
    q^2\phi^2(r)>\frac{(d-3)^2}{4}
\end{equation}
Replace the value of $\phi(r)$ in the confinement phase we have
\begin{equation}
    q\mu>\frac{d-3}{2}
\end{equation}
We obtain
\begin{equation}
    N_c<\frac{4\mu}{d-3}
\end{equation}
With $d=4$, (in this case, we still have the 3 dimension boundary because we have the scale $R_y$) from the instability condition we can study the CSC phase with the maximum number of color $N_{c\text{max(4)}}=6$. With $d=5$ if $\mu_{cd(5)}>1.5$, we have $N_{c\text{max(5)}}=3$ and if $\mu_{cd(5)}\leq 1.5$ the $N_{c\text{max(5)}}=2$. In the $d=6$ case, in the confinement phase $N_{c\text{max(6)}}=2$ but from the estimate in\cite{Kazuo2019}, we do not have the CSC phase in the confinement phase if we only study by Einstein$-$Maxwell gravity. 

We introduce the variable $z=\frac{r_0}{r}$ and replace it in the second equation of \eqref{confinement eom} and set $r_0=1$, we have $f(z)=1-z^{d-1}$. After some manipulation which analogy deconfinement case we obtain
\begin{equation}
    H''(z)+p(z)H'(z)+q(z)H(z)+\lambda^2w(z)\xi(z)^2H(z)=0
\end{equation}
where
\begin{equation}
    \begin{split}
        p(z)&=\frac{f'(z)}{f(z)}+\frac{d-2}{z}\\
        q(z)&=\frac{d-2}{z}\left(\frac{f'(z)}{f(z)}-\frac{1}{z}\right)-\frac{m^2}{f(z)z^2}\\
        &=-\left[\frac{d-2}{z}\frac{(d-2)z^{d-1}+1}{z(1-z^{d-1})}+\frac{m^2}{z^2(1-z^{d-1})}\right]\\
        &=-\frac{(d-2)^2z^{d-1}+(d-2)+m^2}{z^2(1-z^{d-1})}\\
        w(z)&=\frac{1}{f(z)}\\
        \xi(z)&=1
    \end{split}
\end{equation}
And we have the Sturm$-$Liouville form \eqref{Sturm-Liouville equation} of this equation of motion is 
\begin{equation}
    \begin{split}
        Q(z)&=-T(z)q(z)\\
        P(z)&=T(z)w(z)\xi^2(z)\\
        T(z)&=e^{\int p(z)dz}=f(z)z^{d-2}\\
        &=[1-z^{d-1}]z^{d-2}
    \end{split}
\end{equation}
The eigenvalue of this equation is calculated by \eqref{eigenvalue} with the simplest trial function $H(z)=1-az^2$

\begin{equation}
\label{S-L confinement}
    \lambda^2=\frac{\int_0^1 T(z)H'^2(z)dz+\int^1_0 Q(z)H(z)^2dz}{\int^1_0 P(z)H^2(z)dz}
\end{equation}
We can see that in confinement phase the RHS in \eqref{S-L confinement} doesn't depend on the chemical potential.

By solving equation \eqref{S-L confinement} with $H(z)=1-az^2$ and $d=4$ we do not see the chemical potential in $[0,\sqrt{8/3}]$ for the CSC phase transition to occur even with $N_c=1$ (see Fig. \ref{fig02} B). Hence in confinement phase the color superconductivity does not exist with our trial function $H(z)=1-az^2$. But this is also one trial function; we cannot conclude that CSC does not exist. If we want to have CSC phase, we need the other form of the trial function $H(z)$ which still satisfies condition $H(0)=1$ and $H'(0)=0$, now we find the condition of $H(z)$ from $N_c=1$ to $N_c=4$.

To obtain the CSC phase with $N_c=1$ in the confinement phase and $d=4$, we need the trial function $H(z)$ that satisfies $H(0)=1$, $H'(0)=0$ (here we consider that the chemical potential for the full confinement case is not larger than $\sqrt{8/3}$ ) and 
\begin{equation}
\label{condition 4d confinement}
    \text{RHS}<\frac{4}{N_c^2}\times \frac{8}{3},
\end{equation}
with $N_c=1,2,3,4,5,6$. If there is one of $H(z)$ that satisfies these conditions, we have the CSC phase in the confinement phase with this value of $N_c$. If not, we have no CSC in the confinement phase. 

And with the arbitrary $d-$dimension in confinement phase, we obtain the following condition:
\begin{equation}
\label{condition d-dimension confinement}
    \text{RHS}<\frac{4}{N_c^2}\times (\mu_{cd(d)})^2,
\end{equation}
where $\mu_{cd(d)}$is the critical chemical potential of the confinement deconfiment transition correspond to $d-$dimension and $N_c<N_{c\text{max(d)}}$. If we can find (or at least prove that there exists) the function $H(z)$ that satisfies condition \eqref{condition d-dimension confinement} and our boundary condition $H(0)=1$, $H'(0)=0$ for one $N_c<N_c{\text{max(d)}}$, we can study the CSC phase with this number of color by the "pure" Einstein$-$Maxwell gravity. However, because the RHS of \eqref{S-L confinement} does not depend on the chemical potential, the probability that the trial function $H(z)$ exists is very difficult, more difficult than the 4d without confinement case. 

Here, there is one question. For example, in $d=4$ what happens if there exists a $H(z)$ which satisfies the condition for $N_c\geq 3$ in the case without confinement phase or $N_c\geq 7$ in the case of confinement phase? The answer is the instability condition, this condition does not depend on $H(z)$, this is right for all $H(z)$. Hence, with all $H(z)$, we only have the CSC phase with $N_c=2$ in the case without confinement, and $N_c=6$ in the confinement case in $d=4$ if we use the Einstein$-$Maxwell gravity.

\ \


\section{Discussion \label{sec4}}
  
By this holographic model for color superconductivity in the general dimension bulk and general $SU(N_c)$ gauge theory on the boundary, we have found that if we only use Einstein-Maxwell gravity and the standard Maxwell interaction even when we consider the dimension $d=4$ it is very difficult to study the CSC phase with $N_c=2$ (in the higher dimensions bulk there is no CSC phase transition with $N_c=2$ in the deconfinement phase \cite{Vu2024}). In this model, with $d=4$, there is no confinement$-$deconfiement phase transition in case of $T=0$ and $\mu\neq 0$ and the maximum of $\mu$ ($r_+=1$) is $\sqrt{8/3}$. When we study the possibility of the $N_c=2$ CSC phase, we see that in the case without confinement phase with $H(z)=1-az^2$ it does not have $\mu_c$ solution, in the full confinement case and the same trial function there is no solution $\mu_c$ even with $N_c=1$. The critical chemical potential $N_c=2$ for $N_c=2$ CSC only exits if we find a trial function $H(z)$ that satisfies $H(0)=1$, $H'(0)=0$ and the condition \eqref{condition 4d without confinement} for the case without confinement phase and \eqref{condition 4d confinement} for the confinement phase (or at least we prove that these $H(z)$ that satisfy these conditions exist). This means that if there exists color superconductivity for $N_c=2$ then the form of $H(z)$ that satisfies \eqref{condition 4d without confinement} will be very different from $1-az^2$. We hope that we will prove that the function $H(z)$ for $N_c=2$ exists to reach an exact conclusion of this problem in the case of non-confinement. But with $N_c=1$ in the case without confinement with the simplest form of the trial function $H(z)=1-az^2$ we obtained the CSC phase. And from \cite{Vu2024} we cannot study the CSC phase in the 2d boundary (except the case of full confinement in this paper because the number of dimensions of the boundary is still three) because there are no \cite{Wagner1966},\cite{Coleman1973}. If we want to study the CSC phase with $N_c=3$ in the deconfinement or without confinement phase, we must modify gravity \cite{Nam2021} or the Maxwell interaction \cite{Nam2022}. This project doesn't study the energy gap and the back reaction of the Cooper pair yet, and we hope to return to study the gap and the back reaction in the CSC phase via holography earliest. With the other numbers of dimensions, in the deconfinement phase, the Einstein-Maxwell have only CSC phase with $N_c=1$, and in the confinement phase the $N_{cmax}$ depend on the value of the critical chemical potential of the confinement-deconfinement phase transition. 

The further challenges include developing holographic models for, e.g. $p-$wave and $d-$wave CSC (analogy to the $p-$wave and $d-$wave metallic superconductivity \cite{Gubser2008}), \cite{Kim2013} for all confinement and deconfinement phases, as well as examining the Josephson junction effect in the CSC phase through holographic QCD, and after we generalized these for an arbitrary $SU(N_c)$. Moreover, the holographic entanglement entropy in color superconductivity is also one of the interesting topics. These aspects are worth exploring in future works.



A compelling avenue for further research involves holographic modeling in fractional dimensions, where the value $d$ continuously varies. As demonstrated in Fig. \ref{fig01}, this framework situates the dual gauge theory in spacetime dimensions below four, opening possibilities for experimental realizations. Advances in fractal lattice design \cite{Kempkes2019Design} highlight the potential of this direction, given the prevalence of fractional$-$dimensional structures in nature \cite{mandelbrot1982fractal}. Exploring such models could be particularly impactful for condensed matter and quantum gravity, where these unconventional geometries may uncover novel phenomena, as seen in fluid dynamics \cite{phan2024vanishing} and soft matter physics \cite{phan2020bacterial}.

\section{Acknowledgements}

We would like to thank Nam H. Cao for his helpful feedback and insightful discussions. We also acknowledge Trung V. Phan for his assistance in generating Fig. \ref{fig01}, Fig. \ref{fig02} and editing this manuscript and thank grateful to Dmitry Voskresensky for his valuable discussions on color superconductivity.

\bibliography{main}
\bibliographystyle{unsrt}
\end{document}